\documentclass[aps,prl,twocolumn,showpacs,floatfix,longbibliography,superscriptaddress]{revtex4-2}
\usepackage{graphicx,amsfonts,amssymb,amsmath}
\usepackage{textcomp} 
\usepackage{esint}
\usepackage[english]{babel}
\usepackage{afterpage}
\usepackage{bbold}
\usepackage{soul}
\usepackage[colorlinks,linktocpage,citecolor=blue,linkcolor=red,urlcolor=blue]{hyperref}

\input cyracc.def

\allowdisplaybreaks

\begin{document}
	
\graphicspath{{./figures_submit/}}

\allowdisplaybreaks
	

\newcommand{\oldnew}[2]{\marginpar{\scriptsize \textcolor{red}{correction}}{\textcolor{red}{#2}}}
\newcommand{\suppressed}[1]{\marginpar{\scriptsize \textcolor{red}{correction}}{\textcolor{red}{\st{#1}}}}
\newcommand{\correction}[1]{\marginpar{\textcolor{red}{\scriptsize #1}}}
\newcommand{\marge}[1]{\marginpar{\scriptsize #1}}
\newcommand{\remarque}[1]{\marginpar{\scriptsize Remarque}{\it [#1]}}

%

\def\rhoeq{\hat\rho_{\rm eq}}

\newcommand{\new}[1]{{\bf #1}}
\newlength{\textlarg}
\newcommand{\redbar}[1]{\textcolor{red}{\st{#1}}} 
\newcommand{\bluebar}[1]{\textcolor{blue}{\st{#1}}} 

\newcommand{\normord}[1]{:\mathrel{#1}:}

\newcommand{\beq}{\begin{equation}}
\newcommand{\eeq}{\end{equation}}
\newcommand{\bfig}{\begin{figure}}
\newcommand{\efig}{\end{figure}}
\newcommand{\bline}{\begin{multline}}
\newcommand{\eline}{\end{multline}}
\newcommand{\bremark}{\begin{quotation} \noindent \small }
\newcommand{\eremark}{\end{quotation}}
\newcommand{\llbrace}{\left\lbrace}  
\newcommand{\rrbrace}{\right\rbrace}
\newcommand{\lbraket}{\left[}
\newcommand{\rbraket}{\right]}
\newcommand{\llangle}{\left\langle}
\newcommand{\rrangle}{\right\rangle} 

\newcommand{\Tr}{{\rm Tr}} 
\newcommand{\tr}{{\rm tr}} 
\newcommand{\sgn}{\,{\rm sgn}} 
\newcommand{\mean}[1]{\langle #1 \rangle}
\newcommand{\commu}[2]{[#1,#2]} 
\newcommand{\bra}[1]{\langle#1|}
\newcommand{\ket}[1]{|#1\rangle}
\newcommand{\braket}[2]{\langle #1|#2\rangle}
\newcommand{\ketbra}[2]{|#1\rangle\langle#2|}
\newcommand{\dbraket}[3]{\langle #1|#2|#3\rangle}
\newcommand{\tens}[1]{\overleftrightarrow{#1}}  
\newcommand{\vac}{|{\rm vac}\rangle} 
\newcommand{\bravac}{\langle{\rm vac}|}
\newcommand{\const}{{\rm const}} 
\newcommand{\unif}{{\rm unif.}} 
\newcommand{\atanh}{\,{\rm atanh}}
\newcommand{\cotanh}{\,{\rm cotanh}}

\newcommand{\ie}{i.e.\xspace}
\newcommand{\iet}{i.e.}
\newcommand{\eg}{e.g.\xspace}
\newcommand{\cc}{{\rm c.c.}} 
\newcommand{\hc}{{\rm h.c.}} 
\newcommand{\etal}{{\it et al. }}
\newcommand\eme{$^{\mbox{\footnotesize ème}}$\xspace}

\newcommand{\jhatbf}{\hat {\textbf \jold}} 
\newcommand{\Jhatbf}{\hat {\textbf \J}} 
\newcommand{\jhat}{\hat {\jmath}} 
\newcommand{\Jhat}{\hat {J}} 
\newcommand{\jbf}{\textbf j}
\newcommand{\Jbf}{\textbf J}

\def\chibf{\boldsymbol{\chi}}
\def\down{\downarrow}
\def\eps{\epsilon}
\def\gam{\gamma} 
\def\alphabf{\boldsymbol{\alpha}}
\def\phibf{\boldsymbol{\phi}}
\def\varphibf{\boldsymbol{\varphi}}
\def\varphibfs{\boldsymbol{\varphi}_<}
\def\varphibfl{\boldsymbol{\varphi}_>}
\def\varphis{\varphi_{<}}
\def\varphil{\varphi_{>}}
\def\psibf{\boldsymbol{\psi}}
\def\thetabf{\boldsymbol{\theta}}
\def\Ome{\Omega}
\def\omeD{{\omega_D}} 
\def\bfOme{\boldsymbol{\Omega}} 
\def\Omebf{\boldsymbol{\Omega}} 
\def\lamb{\lambda}
\def\Lamb{\Lambda}
\def\sig{\sigma}
\def\Sig{\Sigma}
\def\sigp{{\sigma'}} 
\def\bfsig{\boldsymbol{\sigma}} 
\def\sigbf{\boldsymbol{\sigma}} 
\def\bfSig{\boldsymbol{\Sigma}} 
\def\The{\Theta} 
\def\up{\uparrow}

\def\epsk{\epsilon_{\bf k}} 
\def\epsp{\epsilon_{\bf p}} 
\def\xik{\xi_{\bf k}} 
\def\txik{\tilde\xi_{\bf k}} 
\def\xip{\xi_{\bf p}} 
\def\epsq{\epsilon_{\bf q}} 
\def\xiq{\xi_{\bf q}} 
\def\xikq{\xi_{{\bf k}+{\bf q}}} 
\def\Ek{E_{\bf k}} 
\def\Ep{E_{\bf p}}
\def\Eq{E_{\bf q}}
\def\Heff{\hat H_{\rm eff}}
\def\Hem{\hat H_{\rm em}}
\def\Hint{\hat H_{\rm int}}
\def\Hloc{\hat H_{\rm loc}}
\def\HMF{\hat H_{\rm MF}}
\def\HLL{\hat H_{\rm LL}}
\def\Hdis{\hat H_{\rm dis}}
\def\Sem{S_{\rm em}}
\def\SMF{S_{\rm MF}} 
\def\SHF{S_{\rm HF}} 
\def\SRPA{S_{\rm RPA}} 
\def\Sint{S_{\rm int}} 
\def\Sloc{S_{\rm loc}}
\def\TN{T_{\rm N}} 
\def\TNHF{T^{\rm HF}_{\rm N}} 
\def\Zloc{Z_{\rm loc}} 
\def\ZMF{Z_{\rm MF}} 
\def\ZHF{Z_{\rm HF}} 
\def\ZRPA{Z_{\rm RPA}} 
\def\RPA{{\rm RPA}}
\def\loc{{\rm loc}} 
\def\pp{{\rm pp}}
\def\ph{{\rm ph}} 
\def\ch{{\rm ch}}
\def\sp{{\rm sp}} 
\def\qtf{q_{\rm TF}}
\def\epstf{\eps^{}_{\rm TF}} 
\def\epsrpa{\eps^{}_{\rm RPA}} 
\def\chinnzpp{\chi_{nn}^{0}{}\!\!\!''}
\def\SigHF{\Sigma_{\rm HF}}
\def\psicl{\psi_{\rm cl}} 

\def\half{\frac{1}{2}}
\def\dhalf{\dfrac{1}{2}}
\def\third{\frac{1}{3}} 
\def\quarter{\frac{1}{4}}

\def\qr{{\bf q}\cdot{\bf r}}
\def\wt{\omega t} 

\def\a{{\bf a}}
\def\b{{\bf b}}
\newcommand{\cv}{{\bf c}} 
\def\e{{\bf e}}
\def\f{{\bf f}}
\def\g{{\bf g}}
\def\h{{\bf h}}
\def\jold{\char"11}
\def\j{{\bf j}}
\def\k{{\bf k}}
\def\l{{\bf l}}
\def\ellbf{\bm{\ell}} 
\def\m{{\bf m}}
\def\n{{\bf n}} 
\def\p{{\bf p}} 
\def\q{{\bf q}}
\def\r{{\bf r}}
\def\t{{\bf t}}
\def\u{{\bf u}}
\newcommand{\vv}{{\bf v}}
\def\x{{\bf x}}
\def\y{{\bf y}} 
\def\z{{\bf z}} 
\def\A{{\bf A}}
\def\B{{\bf B}}
\def\D{{\bf D}} 
\def\E{{\bf E}} 
\def\F{{\bf F}} 
\def\H{{\bf H}}  
\def\J{{\bf J}}
\def\K{{\bf K}} 

\def\G{{\bf G}}
\def\L{{\bf L}}
\def\M{{\bf M}}  
\def\O{{\bf O}} 
\def\P{{\bf P}} 
\def\Q{{\bf Q}} 
\def\R{{\bf R}}
\def\S{{\bf S}}
\def\U{{\bf U}} 
\def\V{{\bf V}} 
\def\X{{\bf X}} 
\def\Y{{\bf Y}} 
\def\epsbf{\boldsymbol{\epsilon}}
\def\betabf{\boldsymbol{\beta}}
\def\deltabf{\boldsymbol{\delta}}
\def\mubf{\boldsymbol{\mu}}
\def\nablabf{\boldsymbol{\nabla}}
\def\rhobf{\boldsymbol{\rho}}
\def\sigmabf{\boldsymbol{\sigma}} 
\def\Pibf{\boldsymbol{\Pi}}
\def\pibf{\boldsymbol{\pi}}

\def\para{\parallel}
\def\kpara{{k_\parallel}}
\def\kperp{{k_\perp}} 
\def\kperpp{{k_\perp'}} 
\def\qperp{{q_\perp}} 
\def\tperp{{t_\perp}} 

\def\w{\omega}
\def\wn{\omega_n}
\def\wm{\omega_m}
\def\wnu{\omega_\nu}
\def\wp{\omega_p} 
\def\dmu{{\partial_\mu}}
\def\dnu{{\partial_\nu}}
\def\dl{{\partial_l}}  
\def\dt{\partial_t} 
\def\tdt{\tilde\partial_t}
\def\dk{\partial_k}
\def\tdk{\tilde\partial_k}
\def\dx{\partial_x}
\def\dy{\partial_y} 
\def\dw{\partial_{\w}}
\def\dtau{{\partial_\tau}}  
\def\det{{\rm det}} 
\def\Pf{{\rm Pf}}
\def\diag{{\rm diag}}

\def\dsum{\displaystyle \sum}
\def\dint{\displaystyle \int} 
\def\intt{\int_{-\infty}^\infty dt} 
\def\inttp{\int_{-\infty}^\infty dt'} 
\def\intk{\int_{\bf k}} 
\def\intkd{\int \frac{d^dk}{(2\pi)^d}}
\def\intq{\int_{\bf q}} 
\def\intr{\int d^dr}  
\def\dintr{\displaystyle \int d^dr} 
\def\intrp{\int d^dr'}
\def\dinttau{\displaystyle \int_0^\beta d\tau}
\def\dinttaup{\displaystyle \int_0^\beta d\tau'}
\def\inttau{\int_0^\beta d\tau}
\def\inttaup{\int_0^\beta d\tau'}
\def\intx{\int d^{d+1}x} 
\def\inttaur{\int_0^\beta d\tau \int d^dr}
\def\intinf{\int_{-\infty}^\infty}
\def\dinttaur{\displaystyle \int_0^\beta d\tau \int d^dr}
\def\dintinf{\displaystyle \int_{-\infty}^\infty}
\def\intw{\int_{-\infty}^\infty \frac{d\w}{2\pi}}
\def\sumr{\sum_{\bf r}} 

\def\calA{{\cal A}}
\def\calAbf{\bm{{\cal A}}}
\def\calB{{\cal B}} 
\def\calC{{\cal C}} 
\def\dt{\partial_t}
\def\calD{{\cal D}}
\def\calE{{\cal E}}
\def\calF{{\cal F}} 
\def\calFbf{\bm{{\cal F}}}
\def\calG{{\cal G}}
\def\calH{{\cal H}}
\def\calI{{\cal I}}
\def\calJ{{\cal J}}
\def\calK{{\cal K}}
\def\calL{{\cal L}} 
\def\calM{{\cal M}} 
\def\calN{{\cal N}}
\def\calO{{\cal O}}
\def\calP{{\cal P}}  
\def\calR{{\cal R}} 
\def\calS{{\cal S}}
\def\calT{{\cal T}}
\def\calU{{\cal U}}
\def\calV{{\cal V}}
\def\calX{{\cal X}} 
\def\calY{{\cal Y}} 
\def\calW{{\cal W}} 
\def\calZ{{\cal Z}}

\def\tT{{\tilde T}}
\def\talpha{{\tilde\alpha}}
\def\tbeta{{\tilde\beta}}
\def\tchi{{\tilde\chi}}
\def\tdelta{{\tilde\delta}}
\def\tDelta{{\tilde\Delta}}
\def\teta{{\tilde\eta}} 
\def\tlamb{{\tilde\lambda}}
\def\tmu{{\tilde\mu}}
\def\tphibf{{\tilde\phibf}}
\def\trho{{\tilde\rho}}
\def\tvarphibf{{\tilde\varphibf}} 
\def\tq{\tilde q}
\def\tw{{\tilde\omega}}
\def\twn{{\tilde\omega_n}}
\def\twnu{{\tilde\omega_\nu}}

\def\asinh{{\rm asinh}} 
\def\Tbkt{T_{\rm BKT}}

\title{Mott-glass phase induced by long-range correlated disorder \\ in a one-dimensional Bose gas}

\author{Nicolas Dupuis}
\affiliation{Sorbonne Universit\'e, CNRS, Laboratoire de Physique Th\'eorique de la Mati\`ere Condens\'ee, LPTMC, F-75005 Paris, France}
	
\author{Andrei A. Fedorenko} 
\affiliation{Univ Lyon, ENS de Lyon, CNRS, Laboratoire de Physique, F-69342 Lyon, France} 	
	
\date{October 7, 2024} 
	
\begin{abstract}
We determine the phase diagram of a one-dimensional Bose gas in the presence of disorder with short- and long-range correlations, the latter decaying with distance as $1/|x|^{1+\sig}$. When $\sig<0$, the Berezinskii-Kosterlitz-Thouless transition between the superfluid and the localized phase is driven by the long-range correlations and the Luttinger parameter $K$ takes the critical value $K_c(\sig)=3/2-\sig/2$. The localized phase is a Bose glass for $\sig>\sig_c=3-\pi^2/3\simeq -0.289868$, and a Mott glass ---characterized by a vanishing compressibility and a gapless conductivity--- when $\sig<\sig_c$. Our conclusions, based on the nonperturbative functional renormalization group and perturbative renormalization group, are confirmed by the study of the case $\sig=-1$, corresponding to a perfectly correlated disorder in space, where the model is exactly solvable in the semiclassical limit $K\to 0^+$. 
\end{abstract}
	
\maketitle
	
\paragraph{Introduction.} 

Long-range interactions, decaying as a power law with distance, have crucial consequences on the physical properties of condensed-matter systems. They can alter the phase diagram, the critical properties of phase transitions and the dynamical behavior of physical observables~\cite{Defenu23}. In disordered systems, long-range correlations of the defects play a similar role since they lead, after averaging over the disorder, to effective long-range (elastic) interactions. The modification of the critical behavior and the stabilization of new phases by a long-range correlated disorder has been studied in many systems, including spin systems with random-bond \cite{Dorogovtsev80,Boyanovsky82,Weinrib83,BallesterosParisi:1999,DudkaFedorenkoBlavatskaHolovatch:2016,ChippariPiccoSantachiara:2023} and random-field~\cite{FedorenkoKuhnel:2007,AhrensHartmann:2011} disorder,  different Anderson insulators~\cite{MouraLyra:1998,KuhlIzrailevKrokhin:2008,BillyEtAl:2008}, topological nodal semimetals \cite{Khveshchenko:2007,FedorenkoCarpentierOrignac:2012,LouvetCarpentierFedorenko:2017}, and random elastic manifolds such as domain walls in ferromagnets and flux lines in superconductors~\cite{Balents:1993,Fedorenko06,PetkovicEmigNattermann:2009,Fedorenko:2008,SanchezEtAl:2020,NeverovEtAl:2022}.

\begin{figure}[b]
	\centerline{\includegraphics[width=7cm]{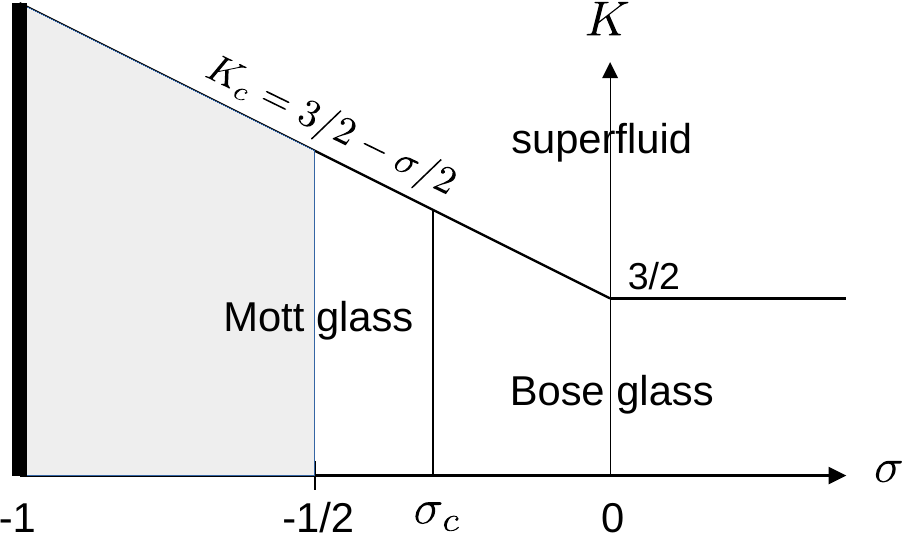}}
	\caption{Phase diagram of a one-dimensional Bose gas in the presence of a (weak) disorder with long-range correlations decaying with distance as $1/|x|^{1+\sig}$. The critical value separating the Bose glass from the Mott glass is $\sig_c=3-\pi^2/3\simeq -0.289868$. The gray area shows the region where the FRG approach [Eqs.~(\ref{rgeq2a}-\ref{rgeq2b})] breaks down. The thick line at $\sig=-1$ corresponds to the case of a perfectly correlated disorder, where the exact solution in the semiclassical limit $K\to 0^+$ confirms that the ground state is a Mott glass 
 } 
	\label{fig:phase_dia}
\end{figure}

In this Letter, we study the phase diagram of a one-dimensional Bose gas in the presence of short- and long-range disorder correlations, the latter decaying with distance as $1/|x|^{1+\sig}$, using bosonization, the replica formalism and the nonperturbative functional renormalization group (FRG) (similar results are obtained with the perturbative field-theoretical FRG~\cite{SM}). 
Our main results are summarized in Fig.~\ref{fig:phase_dia}. The  Berezinskii-Kosterlitz-Thouless transition between the superfluid and the localized phase, which occurs for $K_c=3/2$ (for weak disorder) in the absence of long-range correlations, occurs for $K_c(\sig)=3/2-\sig/2$ when the exponent $\sig$ is negative. The long-range disorder correlations are nevertheless irrelevant in the localized phase when $\sig>\sig_c=3-\pi^2/3\simeq -0.289868$ and the system is a Bose glass with a nonzero compressibility and a gapless conductivity~\cite{Dupuis19,Dupuis20}. The ``short-range'' fixed point is unstable when $\sig<\sig_c$. We find that the localized phase is described by a ``long-range'' fixed point, corresponding to a Mott glass with a vanishing compressibility and a gapless conductivity, but point out that the cumulant expansion used in the RG approach formally breaks down. In the case $\sig=-1$, corresponding to a perfectly correlated disorder in space, the model in exactly solvable in the semiclassical limit $K\to 0^+$ and we find a vanishing compressibility and a gapless conductivity varying at $|\w|^3$ in the small-frequency limit. While this result strongly supports the existence of a Mott glass for $\sig<\sig_c$, we cannot exclude a quantum phase transition at a critical value $-1\leq \sig^*<\sig_c$ between two incompressible localized phases with different low-frequency conductivity (e.g. $\w^2$ vs $|\w|^3$).

\paragraph{Model and FRG formalism.}

We consider a one-dimensional Bose gas described by the Luttinger-liquid Hamiltonian
\beq
\HLL = \int dx \frac{v}{2\pi} \left[ \frac{1}{K} (\dx\hat\varphi)^2 + K (\dx\theta)^2 \right] 
\eeq 
in the bosonization formalism~\cite{Giamarchi_book}, where $K$ is the Luttinger parameter and $v$ the velocity of the sound mode. The operator $\hat\theta$ is the phase of the boson operator $\hat\psi=e^{i\theta}\sqrt{\hat\rho}$ and $\dx\hat\varphi$ is conjugate to $\hat\theta(x)$: $[\hat\theta(x),\partial_y\hat\varphi(y)]=i\pi\delta(x-y)$. A random potential that couples to the boson density is taken into account by the Hamiltonian   
\beq 
\Hdis = \int dx \left[ - \frac{1}{\pi} \eta \dx \varphi + ( \rho_2 \xi^* e^{2i\hat\varphi} + {\rm H.c.}) \right] , 
\eeq
where $\eta(x)$ (real) and $\xi(x)$ (complex) denote random potentials with Fourier components near 0 and $\pm 2\pi\rho_0$, respectively~\cite{Giamarchi87,Giamarchi88}. $\rho_0$ is the average particle density and $\rho_2$ a nonuniversal parameter that depends on microscopic details. $\eta(x)$ can be eliminated by a shift of $\varphi$ and is not considered in the following. The random potential $\xi(x)$ is assumed to be Gaussian correlated with zero mean and variance
\beq
\begin{gathered}
	\overline{\xi(x)\xi(x')} = \overline{\xi^*(x)\xi^*(x')} = 0 , \\ 
	\overline{\xi^*(x) \xi(x')} = D_{\rm SR} \delta(x-x') + D_{\rm LR} f(x-x') ,
\end{gathered}
\label{Ddef}
\eeq 
where the overbar denotes the average over disorder. The delta correlated part corresponds to short-range correlations while $f(x)\sim 1/|x|^{1+\sig}$ describes long-range correlations. When $\sig>0$, the Fourier transform $f(q)$ of the function $f(x)$ has a well-defined limit when $q\to 0$ and $f(x)$ is effectively short-range, merely renormalizing the short-range part described by $D_{\rm SR}$. In the following we assume $\sig\leq 0$ so that $f(q)$ diverges when $q\to 0$.  We normalize $f(x)$ so that $f(q)=|q|^{\sig}$ for $q\to 0$ if $\sig<0$ and $f(q\to 0)=\ln(1/|q|)$ if $\sig=0$. One can set $D_{\rm SR}=0$ if the disorder has no short-range correlations, however a nonzero $D_{\rm SR}$ is always generated under coarse graining~\cite{Fedorenko06}. 

We average over disorder by introducing $n$ replicas of the system~\cite{Giamarchi88,Dupuis20,Tarjus19}. After integrating out the phase field $\theta$, we obtain the Euclidean action 
\begin{multline} 
	S[\{\varphi_a\}] = \sum_a \int_{x,\tau} \frac{v}{2\pi K} \left[ (\dx\varphi_a)^2 + \frac{(\dtau\varphi_a)^2}{v^2} \right]  \\
	- \sum_{a,b} \int_{x,x',\tau,\tau'} [ \calD_{\rm SR} \delta(x-x') \\ + \calD_{\rm LR} f(x-x') ]  \cos[2\varphi_a(x,\tau) - 2 \varphi_b(x',\tau') ] ,
	\label{action}
\end{multline}
where $a,b=1,\dots,n$ are replica indices, $\calD_{\rm SR}=\rho_2^2 D_{\rm SR}$, $\calD_{\rm LR}=\rho_2^2 D_{\rm LR}$ and we use the notation $\int_{x,\tau}=\inttau\int dx$. We consider only the zero-temperature limit $\beta=1/T\to\infty$. Note that the disorder is correlated at long distances in (imaginary) time, hence the nonlocal nature of the two-replica potential even for short-range correlations in space. 

\begin{figure*}
	\centerline{\includegraphics[width=8.cm]{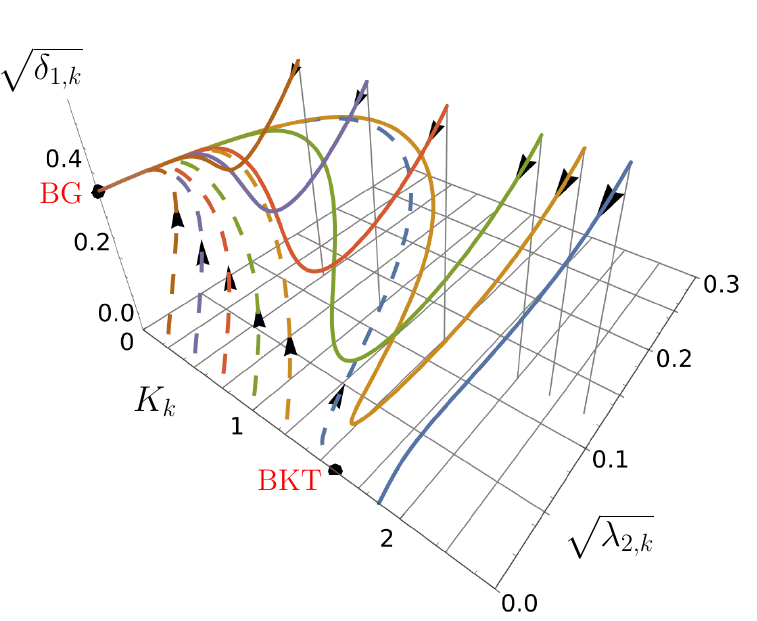}
		\includegraphics[width=8.cm]{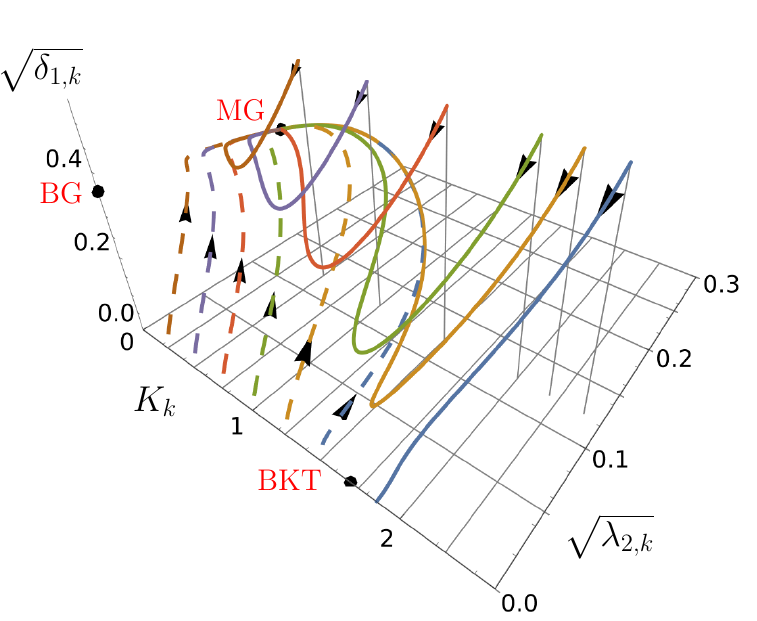}}
	\caption{Flow diagrams in the space $(K,\sqrt{\lambda_2},\sqrt{\delta_1})$ for $\sig=-0.2$ (left) and $\sig=-0.4$ (right). The black dots show the BKT fixed point $((3-\sig)/2,0,0)$ controlling the transition between the superfluid phase and the localized phase, as well as the Bose-glass (BG) and Mott-glass (MG) fixed points. (The dashing of some lines and the different colors are used to improve visibility and have no particular physical meaning.)} 
	\label{fig_flow_diag} 
\end{figure*}

Most physical quantities can be obtained from the effective action (or Gibbs free energy),
\beq 
\Gamma[\{\phi_a\}] = - \ln Z[\{J_a\}] + \sum_a \int_{x,\tau} J_a \phi_a ,
\eeq 
defined as the Legendre transform of the free energy $-\ln Z[\{J_a\}]$. Here $J_a$ is an external source which couples linearly to the field $\varphi_a$ and allows us to obtain the expectation value $\phi_a(x,\tau)=\mean{\varphi_a(x,\tau)}=\delta \ln Z[\{J_a\}]/\delta J_a(x,\tau)$. We compute $\Gamma[\{\phi_a\}]$ using a nonperturbative FRG approach~\cite{Berges02,Delamotte12,Dupuis_review}. In practice, we consider a scale-dependent effective action $\Gamma_k[\{\phi_a\}]$ which includes fluctuations with momenta (and frequencies) between the UV cutoff $\Lambda$ of the theory and a running momentum scale $k\leq\Lambda$. The effective action $\Gamma_{k=0}[\{\phi_a\}]$ of the original model is obtained when all fluctuations have been integrated out whereas $\Gamma_\Lambda[\{\phi_a\}]=S[\{\phi_a\}]$ is given by the microscopic action. $\Gamma_k$ satisfies a flow equation which allows us to obtain $\Gamma_{k=0}$ from $\Gamma_\Lambda$ but which cannot be solved exactly~\cite{Wetterich93,Ellwanger94,Morris94}. We refer to the Supplemental Material~\cite{SM} for a more detailed presentation of the FRG approach as well as a discussion of the perturbative field-theoretical approach.  

We consider the following truncation of the effective action, 
\beq 
\Gamma_k[\{\phi_a\}] = \sum_a \Gamma_{1,k}[\phi_a] - \half \sum_{a,b} \Gamma_{2,k}[\phi_a,\phi_b] ,  
\label{ansatz1} 
\eeq 
with the ansatz 
\begin{equation}
	\Gamma_{1,k}[\phi_a] = \int_{x,\tau} \frac{v_k}{2\pi K_k} \left[ (\dx\phi_a)^2 + \frac{(\dtau\phi_a)^2}{v_k^2} \right]
	\label{ansatz2a} 
\end{equation} 
and  
\begin{multline} 
	\!\!\!\!\!  \Gamma_{2,k}[\phi_a,\phi_b] = \int_{x,x',\tau,\tau'} \bigl[ \delta(x-x') V_{1,k}(\phi_a(x,\tau)-\phi_b(x',\tau')) \\ + f(x-x') V_{2,k}(\phi_a(x,\tau)-\phi_b(x',\tau')) \bigr] .
	\label{ansatz2b} 
\end{multline} 
Equation~(\ref{ansatz1}) is based on a free-replica-sum expansion where all ``cumulants'' $\Gamma_{i,k}$ with $i\geq 3$ are ignored. The form of $\Gamma_{1,k}$ follows from a derivative expansion to second order. The coefficients of the derivative terms are written in terms of a Luttinger parameter $K_k$ and a velocity $v_k$ which are scale dependent. This defines a scale-dependent superfluid stiffness $\rho_{s,k} = v_kK_k/\pi$ and a scale-dependent compressibility $\kappa_k = K_k/\pi v_k$.  The two-replica part $\Gamma_{2,k}$ is assumed to keep the same form as the two-replica term in the microscopic action but with renormalized potentials $V_{1,k}$ and $V_{2,k}$. The initial conditions are 
$K_\Lambda=K$, $v_\Lambda = v$, $V_{1,\Lambda}(u)= 2 \calD_{\rm SR} \cos(2u)$ and $V_{2,\Lambda}(u) = 2 \calD_{\rm LR} \cos(2u)$. 

The flow equations are obtained by inserting the ansatz~(\ref{ansatz1}-\ref{ansatz2b}) into the exact flow equation satisfied by the effective action, 
\begin{align}
	k\dk \Delta_1 ={}& (-3+2\theta-2z+2) \Delta_1 - K l_1 \Delta_1'' \nonumber\\ & + \bar L_0 [ \Delta_1'{}^2 + \Delta_1''(\Delta_1-\Delta(0))]  \nonumber\\ & + \bar L_1 [2\Delta_1 ' \Delta_2' + \Delta_1''(\Delta_2-\Delta_2(0)) + \Delta_1 \Delta_2''] \nonumber\\ & + \bar L_2 (\Delta_2'{}^2 + \Delta_2 \Delta_2'') , \label{rgeq2a} \\ 
	k\dk \Delta_2 ={}& (-3+\sig+2\theta-2z+2) \Delta_2 - K l_1 \Delta_2'' \nonumber\\ & - \bar L_0 \Delta_2'' \Delta_1(0)  - \bar L_1 \Delta_2'' \Delta_2(0) 
	\label{rgeq2b}
\end{align}
(we have dropped the $k$ index in $\Delta_i$, $K$, $\theta$, and $z$ for ease of notations), where 
\beq 
\begin{split}
	\Delta_{1,k}(u) &= - \frac{K_k^2 V''_{1,k}(u)}{v_k^2 k^3} ,  \;\;\;
	\Delta_{2,k}(u) = - \frac{K_k^2 V''_{2,k}(u)}{v_k^2 k^{3-\sig}} ,
\end{split}
\eeq 
are dimensionless potentials, and $\theta_k= k\dk \ln K_k$, $z_k-1= k\dk \ln v_k$ ($z_k$ is the dynamical exponent),
\begin{align}
	\theta_k 
	&= \half [ \bar M_\tau \Delta_{1,k}''(0) + (\bar L_x + \bar L_\tau) \Delta_{2,k}''(0)] , \label{rgeq1a} \\ 
	z_k-1 
	&= \half [ \bar M_\tau \Delta_{1,k}''(0)  - ( \bar L_x - \bar L_\tau)  \Delta_{2,k}''(0)] \label{rgeq1b} .
\end{align}
The ``threshold functions'' $\bar L_i$ ($i\in[1,3]$), $\bar L_x$, $\bar L_\tau$ and $\bar M_\tau$ are $k$-independent linear functions of $\theta_k$ and $z_k$~\cite{SM}. In the absence of long-range correlated disorder, the coefficient $v_k/2\pi K_k$ of the space derivative term in $\Gamma_1$ does not renormalize, in agreement with the statistical tilt symmetry of the action~(\ref{action})~\cite{Dupuis20,not1}.

Equations~(\ref{rgeq2a}-\ref{rgeq1b}) can be solved numerically expanding the $\pi$-periodic functions $\Delta_{1,k}(u)=\sum_{n=1}^\infty \delta_{n,k} \cos(2nu)$ and $\Delta_{2,k}(u)= \lambda_{2,k} \cos(2u)$ in circular harmonics (with an upper cutoff $n_{\rm max}<\infty$ on the number harmonics). The function $\Delta_{2,k}(u)$ has a single harmonics when $k=\Lambda$ and this form is obviously preserved by the flow equation~(\ref{rgeq2b}). Figure~\ref{fig_flow_diag} shows RG flows, for $\sig=-0.2$ and $\sig=-0.4$, projected onto the plane $(K,\sqrt{\lambda_2},\sqrt{\delta_1})$. In both cases one observes a critical BKT fixed point controlling a transition between a superfluid phase and a localized phase. For $\sig=-0.2$, the localized phase is described by an attractive fixed point where the long-range correlated disorder vanishes, $\lambda_2^*=0$, and corresponds to the Bose-glass phase described in previous studies~\cite{Dupuis19,Dupuis20a}. For $\sig=-0.4$, the long-range part of the disorder does not vanish at the fixed point describing the localized phase ($\lambda_2^*\neq 0$); we shall see that this fixed point describes a Mott glass.

\paragraph{Superfluid-insulator transition.} 

\begin{figure}[b]
	\centerline{\includegraphics[width=8cm]{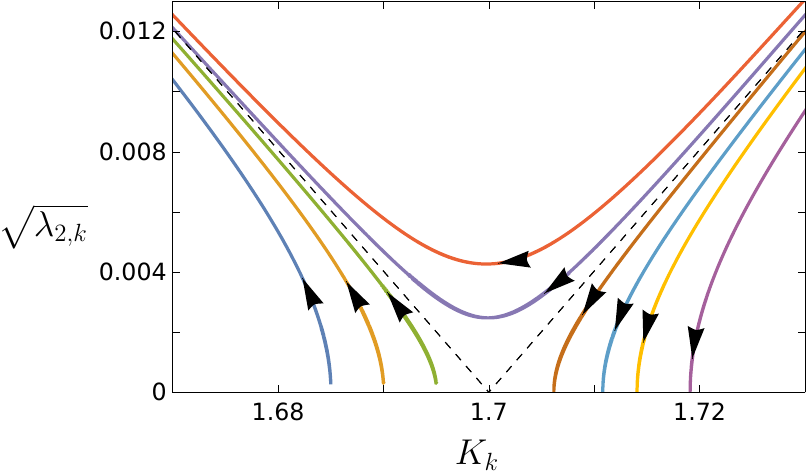}}
	\caption{Flow trajectories, projected in the plane $(K,\sqrt{\lambda_2})$, near the BKT fixed point located at $K_c=1.7$ ($\sig=-0.4$). The initial conditions are defined by $\delta_{1,\Lambda}\sim 0.1$ when $K_\Lambda>1.7$, and $\delta_{1,\Lambda}\sim 10^{-6}$  when $K_\Lambda<1.7$. The dashed lines show the two separatrixes obtained from the standard BKT flow equations derived from~(\ref{rgeq2a}-\ref{rgeq1b}) to leading order in $K_k-K_c(\sig)$ and $\sqrt{\lambda_{2,k}}$~\cite{SM}. The different colors of the trajectories are used to improve visibility and have no particular physical significance.} 
	\label{fig_flowBKT} 
\end{figure}

The stability of the superfluid phase against an infinitesimal disorder can be determined from scaling analysis considering the microscopic action~(\ref{action}). At the Luttinger-liquid fixed point, $e^{2i\varphi}$ has scaling dimension $K$, so that $[D_{\rm SR}]=3-2K$ and $[D_{\rm LR}]=3-\sig-2K$ using $[x]=[\tau]=-1$ and $[f(x)]=1+\sig$. The long-range correlated part of the disorder becomes relevant for $K<K_c(\sig)=3/2-\sig/2$ and $\sig<0$, destabilizing the superfluid phase, so that the critical value of the Luttinger parameter, $K_c(\sig)$, is larger than the value $3/2$ obtained for disorder with only short-range correlations.
By neglecting the short-range correlated disorder, which is irrelevant near the BKT fixed point when $\sig<0$, the flow equations~(\ref{rgeq2a}-\ref{rgeq1b}) reproduce the standard RG equations of the BKT theory to leading order in $K_k-K_c(\sig)$ and $\sqrt{\lambda_{2,k}}$~\cite{SM}. These equations are in very good agreement with the numerical solution of the full equations (Fig.~\ref{fig_flowBKT}). 

\paragraph{Short-range fixed point: Bose glass.} 

In addition to the BKT fixed point, the flow equations~(\ref{rgeq2a}-\ref{rgeq1b}) admit a fully attractive short-range fixed point defined by $K^*=v^*=0$, where the long-range correlated part of the disorder is irrelevant: $\Delta_2^*(u)=0$ (Fig.~\ref{fig_flow_diag}, left panel). Since $\lim_{k\to 0}\theta_k=\theta^*=z^*-1$, which follows from a restoration of the statistical tilt symmetry at the fixed point, the superfluid stiffness $\rho_{s,k} = v_kK_k/\pi \sim k^{\theta^*+z^*-1}\to 0$ vanishes at the fixed point, but the compressibility $\kappa_k = K_k/\pi v_k \sim  k^{\theta^*-z^*+1}\to\const$ remains finite. The system is not superfluid but compressible, as expected for a Bose glass.

\begin{figure}
	\centerline{\includegraphics[width=7cm]{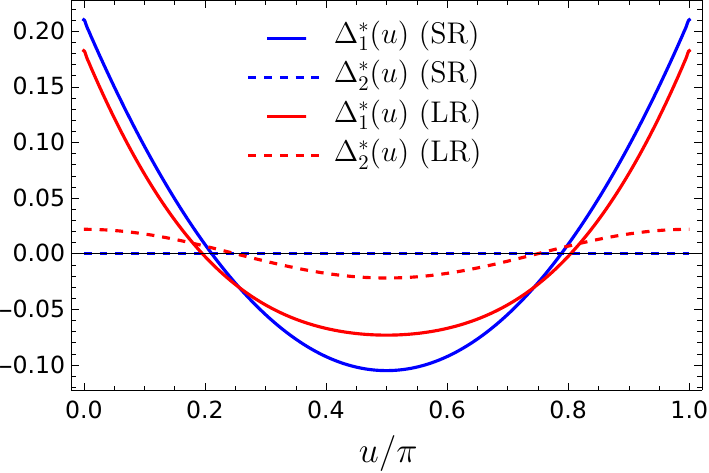}}
	\caption{Disorder correlators $\Delta_1^*(u)$ and $\Delta_2^*(u)$ at the short-range (SR) and long-range (LR) fixed points describing the Bose glass and the Mott glass, respectively.}
	\label{fig_pot} 
\end{figure}

The value of $\Delta_ 1^*(u)$ can be found analytically~\cite{Dupuis20}, 
\beq 
\Delta_ 1^*(u) = \frac{1}{2\bar L_0} \left[ \left(u-\frac{\pi}{2}\right)^2 - \frac{\pi^2}{12} \right] , 
\label{Delta1SRFP} 
\eeq 
and exhibits ``cusps'' at $u=0,\pm\pi,\pm 2\pi,\dots$ (Fig.~\ref{fig_pot}). A cuspy functional form of the renormalized disorder correlator is known to be related to the existence of metastable states and glassy properties~\cite{Balents96}. At nonzero momentum scale $k$, quantum tunneling between the ground state and these metastable states leads to a rounding of the cusp singularity into a quantum boundary layer. The latter controls the low-energy dynamics and is responsible for the $\sig(\w)\sim \w^2$ behavior of the (dissipative) conductivity. We refer to Refs.~\cite{Dupuis19,Dupuis20,Daviet21} for a detailed discussion.

The stability of the short-range fixed point with respect to long-range disorder correlations can be determined by considering the linearized equation satisfied by $\Delta_{2,k}(u)$, 
\beq 
k\dk \Delta_{2,k}(u) = (-3+\sig) \Delta_{2,k}(u) - \bar L_0 \Delta_1^*(0) \Delta''_{2,k}(u) . 
\eeq 
Using $\Delta_1^*(0)=\pi^2/12\bar L_0$ , we obtain 
\beq 
k\dk \lambda_{2,k} = \left(-3+\sig+\frac{\pi^2}{3} \right) \lambda_{2,k} . \label{eq:stability}
\eeq 
The Bose-glass fixed point is therefore unstable when $\sig<\sig_c$, where $\sig_c = 3 - \pi^2/3\simeq -0.289868$. 

\paragraph{Long-range fixed point: Mott glass.} 

When $\sig<\sig_c$, we find that the localized phase is described by a fully attractive fixed point where $K^*=v^*=0$ and both disorder correlators $\Delta_1^*(u)$ and $\Delta_2^*(u)$ are nonzero (Fig.~\ref{fig_flow_diag}, right panel). In marked contrast to the short-range fixed point, $\theta^*\neq z^*-1$ (the statistical tilt symmetry is not restored at the fixed point), so that the compressibility $\kappa_k\sim k^{\theta^*-z^*+1}\to 0$ vanishes: The system is incompressible. Although the functions $\Delta_1^*(u)$ and $\Delta_2^*(u)$ cannot be found analytically, they can be obtained from the solution of the flow equations in the limit $k\to 0$ (Fig.~\ref{fig_pot}). $\Delta_1^*(u)$ differs from~(\ref{Delta1SRFP}) but still exhibits cusps at $u=n\pi$ ($n$ integer). This cusp singularity and the quantum boundary layer at nonzero RG scale $k$ describes the physics of rare low-energy metastable states and their coupling to the ground state by quantum fluctuations, as in the Bose-glass phase. Thus the only difference between the short-range and long-range fixed points is the vanishing of the compressibility in the latter case. The long-range fixed point describes an incompressible phase with a gapless conductivity $\sigma(\w)\sim \w^2$, i.e. a Mott glass. 

The FRG approach fails for $\sigma\leq-0.5$ since the threshold function $\bar L_x$ diverges~\cite{SM}. If we include the third cumulant $\Gamma_3$ in the three-replica-sum expansion~(\ref{ansatz1}), we find that some threshold functions diverge for $\sigma\to -1/3$. We conclude that the cumulant expansion of the functional $\Gamma_k[\{\phi_a\}]$ breaks down in the Mott glass. We argue below, considering an exactly solvable limit, that the existence of a Mott glass for all values $\sig<\sig_c$, as suggested by the FRG approach, is nevertheless likely to be correct.

\paragraph{The limit $\sig=-1$ and $K\to 0^+$.} 

When $\sigma=-1$, the random potential $\xi(x)\equiv\xi=|\xi|e^{i\alpha}$ is perfectly correlated in both imaginary-time and space dimensions. For a given realization of the disorder, the action reads
\beq 
S= \int_{x,\tau} \llbrace \frac{v}{2\pi K} \left[ (\dx\varphi)^2 + \frac{(\dtau\varphi)^2}{v^2} \right] - 2|\xi| \cos(2\varphi) \rrbrace
\eeq
after a shift of $\varphi$ to eliminate the phase $\alpha$, and corresponds to a sine-Gordon model with a random coupling constant $|\xi|$. The BKT critical point for this model is located at $K_c=2$, which agrees with the result $K_c(\sig)=3/2-\sig/2$ obtained for an arbitrary $\sig<0$. In the semiclassical limit $K\to 0^+$~\cite{Rajaraman_book}, one can expand the cosine term about one of its minima, e.g. $\varphi=0$, to obtain a Gaussian action, 
\beq 
S = \int_{x,\tau} \frac{1}{2\pi Kv} \bigl[ v^2 (\dx\varphi)^2 + (\dtau\varphi)^2 + m^2 \varphi^2 \bigr] ,
\eeq
where $m=(8\pi vK|\xi|)^{1/2}$ is a random mass. It is now straightforward to compute physical quantities and perform the disorder averaging with the Gaussian probability distribution defined by~(\ref{Ddef})~\cite{SM}. One finds that the compressibility vanishes whereas the conductivity is gapless. The ground state is a Mott glass as suggested by the FRG approach.  Interestingly, for $\sig=-1$ the conductivity vanishes as $|\w|^3$ and not as $\w^2$ as expected in a Bose glass and likely also in a Mott glass. Whether this is a peculiarity of the case $\sig=-1$ ---which could be explained by the fact that a disorder perfectly correlated in space is clearly not self-averaging, thus making the case $\sig=-1$ special--- or holds in the entire Mott glass phase is an open issue; another possibility would be a quantum phase transition at a critical value $-1<\sig^*<\sig_c$ between two Mott-glass phases with different low-frequency conductivities. 

Although the first cumulant $W_1[J_a]=\overline{W[J_a;\xi]}$ of the random functional $W[J;\xi]=\ln Z[J;\xi]$ is well defined ---here $Z[J;\xi]$ is the partition function of the system and $J$ an external source that couples linearly to the field--- we find that the second cumulant $W_2[J_a,J_b]$ and its Legendre transform $\Gamma_2[\phi_a,\phi_b]$ do not exist. We conclude that the cumulant expansion is not defined and there is even no guarantee that their generating functionals $\ln Z[\{J_a\}]$ and $\Gamma[\{\phi_a\}]$ exist. This is in line with the failure of the cumulant expansion observed in the FRG approach when $\sig<\sig_c$ and strongly suggests that the ground state is a Mott glass for all values of $\sig<\sig_c$.

\paragraph{Conclusion.} 

The Mott glass is a very intriguing phase of matter since it is associated with gapped single-particle excitations (which implies a vanishing compressibility) and gapless particle-hole excitations (hence the absence of gap in the conductivity). It has been pointed out that these two properties are compatible only in the presence of long-range interactions~\cite{Nattermann07}. In the present study, long-range (elastic) interactions are a consequence of long-range disorder correlations. 

There have been various proposals for the existence of a Mott glass in one-dimensional quantum fluids. It has been claimed that the interaction between disorder and a commensurable periodic potential can stabilize a Mott glass~\cite{Orignac99,Giamarchi01}, but this proposal does not take into account the necessary long-range interactions~\cite{Nattermann07,Ledoussal08a}. It has also been claimed that the ground state of a disordered system with linearly confining interactions (disordered Schwinger model) is a Mott glass~\cite{Chou18,Giamarchi01} but this conclusion is in contradiction with another study~\cite{Dupuis20a}. The work reported in Ref.~\cite{Daviet20} predicts a Mott glass in a disordered Bose gas with long-range interactions; this conclusion however assumes a ``jellium'' model, where the global neutrality of the system is ensured by a uniform background of opposite charges, which is undoubtedly difficult to achieve experimentally.

Thus we believe that long-range disorder correlations appear as the most natural way proposed so far to observe a Mott glass in a one-dimensional Bose gas. Cold atoms are a natural platform for experimental study. Laser speckles allow to control, to some extent, the disorder potential acting on the gas~\cite{Clement06,Piraud13}, but long-range power-law correlations seem out of reach. On the other hand, by generating the disorder potential from the spatial modulation of a laser beam using a digital micromirror device (DMD), the statistical properties of the disorder potential can be adjusted at will, which could open up the possibility of creating disorder potentials with the desired characteristics (see, e.g., Refs.~\cite{Choi16,RubioAbadal19,Leonard23}). Finally we point out that a classical analog of the Mott glass is expected to exist in heavy-ion-irradiated  type-II superconductors with algebraically correlated columnar disorder~\cite{SM}.

We thank T. Chalopin, D. Cl\'ement and N. Cherroret for useful discussions on the experimental realization of random potentials with long-range correlations in ultracold gases, and E. Orignac for valuable comments.

%
 

\begin{thebibliography}{55}
\bibitem [{not()}]{not1}%
  \BibitemOpen
  \bibinfo {note} {The statistical symmetry is defined by the invariance of the
  disorder part of the action~(\ref{action}) in the change
  $\varphi_a(x,\tau)\to\varphi_a(x,\tau)+w(x)$, where $w(x)$ is an arbitrary
  function of space. It ensures that the coefficient $v_k/\pi K_k$ of the
  space-derivative term in $\Gamma_{1,k}$ is not renormalized.}\BibitemShut
  {Stop}%
\end{thebibliography}

\clearpage

\onecolumngrid


\begin{center}
	\textbf{\large  Mott-glass phase induced by long-range correlated disorder \\ in a one-dimensional Bose gas
		\\ 
		[.3cm] -- Supplemental Material --} \\ 
	[.4cm] Nicolas Dupuis$^1$ and Andrei A. Fedorenko$^2$\\[.1cm]
	{\itshape $^1$Sorbonne Universit\'e, CNRS, Laboratoire de Physique Th\'eorique de la Mati\`ere Condens\'ee, LPTMC, F-75005 Paris, France} \\
	{\itshape $^2$Univ Lyon, ENS de Lyon, CNRS, Laboratoire de Physique, F-69342 Lyon, France}\\
\end{center}

\setcounter{equation}{0}
\setcounter{figure}{0}
\setcounter{table}{0}
\renewcommand{\theequation}{S\arabic{equation}}
\renewcommand{\theHequation}{S\arabic{equation}}
\renewcommand{\thefigure}{S\arabic{figure}}	
\renewcommand{\bibnumfmt}[1]{[S#1]}
\renewcommand{\citenumfont}[1]{S#1}


\bigskip 
\bigskip 

In the Supplemental Material, we describe in more detail the FRG approach discussed in the main text. We then consider the perturbative field-theoretical RG and show that it fully agrees with the predictions of the FRG.  We also analyze the case $\sig=-1$, corresponding to a perfectly correlated disorder in space, which is exactly solvable in the limit $K\to 0^+$. Finally, we discuss the application of our results to the pinning of vortices by correlated disorder in type II superconductors.

\section{I. Nonperturbative functional renormalization group}

The strategy of the nonperturbative FRG approach is to build a family of models indexed by a momentum scale $k$ such that fluctuations are smoothly taken into account as $k$ is lowered from the UV scale $\Lambda$ down to 0. This is achieved by adding to the action of the replicated system the infrared regulator term 
\beq 
\Delta S_k[\{\varphi_a\}] = \half \sum_{a,q,\wn} \varphi_a(-q,-i\wn) R_k(q,i\wn) \varphi_a(q,i\wn) ,
\eeq
where $\wn=2n\pi T$ ($n\in \mathbb{Z}$) is a Matsubara frequency and $a$ a replica index. The cutoff function $R_k(q,i\wn)$ is chosen so that fluctuation modes satisfying $|q|,|\w|/v_k \ll k$ are suppressed while those with $|q|>k$ or $|\w|/v_k>k$ are left unaffected ($v_k$ denotes the $k$-dependent sound-mode velocity, see Eq.~(\ref{ansatz2a}) in the main text). In practice we take the cutoff function 
\beq
R_k(q,i\wn) = \frac{v_k}{\pi K_k} \left( q^2 + \frac{\wn^2}{v_k^2} \right) r\left( \frac{q^2}{k^2} + \frac{\wn^2}{v_k^2 k^2} \right) , 
\label{Rdef}
\eeq 
where $r(y)=\alpha/(e^y-1)$ with $\alpha$ a constant of order unity. 

The partition function
\beq 
\calZ_k[\{J_a\}\}] = \int \calD[\{\varphi_a\}]\, e^{-S[\{\varphi_a\}] - \Delta S_k[\{\varphi_a\}] + \sum_a \int_{x,\tau} J_a\varphi_a } ,
\eeq 
written here in the presence of external sources, becomes $k$ dependent. The expectation value of the field,
\beq 
\phi_{a,k}[x,\tau;\{J_f\}] = \frac{\delta \ln \calZ_k[\{J_f\}]}{\delta J_a(x,\tau)}
= \mean{\varphi_a(x,\tau)} ,
\eeq 
is obtained from a functional derivative with respect to the external source $J_a$ (to avoid confusion in the indices, we denote by $\{J_f\}$ the $n$ external sources). The scale-dependent effective action
\beq 
\Gamma_k[\{\phi_a\}] = - \ln \calZ_k[\{J_a\}] + \sum_a \int_{x,\tau} J_a \phi_a - \Delta S_k[\{\phi_a\}] 
\eeq 
is defined as a modified Legendre transform which includes the subtraction of $\Delta S_k[\{\phi_a\}]$. Assuming that for $k=\Lambda$ the fluctuations are completely frozen by the $\Delta S_k$ term, $\Gamma_\Lambda[\{\phi_a\}]=S[\{\phi_a\}]$. On the other hand, the effective action of the original model~(\ref{action}) is given by $\Gamma_{k=0}$ provided that $R_{k=0}$ vanishes. The nonperturbative FRG approach aims at determining $\Gamma_{k=0}$ from $\Gamma_\Lambda$  using Wetterich’s equation~\cite{Wetterich93sm,Ellwanger94sm,Morris94sm},  
\beq 
\dt\Gamma_k[\{\phi_a\}] = \half \Tr \llbrace \dt R_k \bigl( \Gamma_k^{(2)}[\{\phi_a\}] + R_k \bigr)^{-1} \rrbrace , 
\label{Weteq}
\eeq 
where $\Gamma_k^{(2)}$ is the second functional derivative of $\Gamma_k$ and $t = \ln(k/\Lambda)$ a (negative) RG ``time''. The trace involves a sum over momenta and frequencies as well as the replica index.

\subsection{A. Flow equations} 

Inserting the ansatz~(\ref{ansatz1}-\ref{ansatz2b}) into the Wetterich equation~(\ref{Weteq}), we obtain the flow equations~(\ref{rgeq2a}-\ref{rgeq1b}) of the main text. In the zero-temperature limit, the threshold functions are defined by 
\beq
\begin{split}
	l_1 &= - \pi \tdt \int \frac{d\tilde q}{2\pi} \int \frac{d\tw}{2\pi} \tilde G(\tilde q,i\tw) , \\ 
	\bar L_n &= - \pi^2 \tdt \int \frac{d\tilde q}{2\pi} f(\tq)^n \tilde G(\tq,0)^2 \quad (n=0,1,2), \\  
	\bar L_x &= - \frac{\pi^2}{2} \tdt \int \frac{d\tilde q}{2\pi} f(\tq) \tilde G^{(2,0)}(\tq,0) , \\ 
	\bar L_\tau &= - \frac{\pi^2}{2} \tdt \int \frac{d\tilde q}{2\pi} f(\tq) \tilde G^{(0,2)}(\tq,0) , \\ 
	\bar M_\tau &= - \frac{\pi^2}{2} \tdt \int \frac{d\tilde q}{2\pi}\tilde G^{(0,2)}(\tq,0) ,
\end{split}
\eeq  
where 
\beq 
\tilde G(\tq,i\tw) = \frac{1}{(\tq^2+\tw^2)[1+r(\tq^2+\tw^2)]},
\eeq 
with $\tq=q/k$ and $\tw=\w/v_kk$, 
is the dimensionless propagator and $\tilde G^{(i,j)}(\tq,i\tw)=\partial^i_{\tq} \partial^j_{\tw} \tilde G(\tq,i\tw)$. Here $\tdt=(\dt R_k)\partial_{R_k}$ acts only on the $k$ dependence of the cutoff function $R_k$.

\subsection{B. Superfluid--Bose-glass transition}

Near the BKT point defined by $K^*=3/2-\sig/2$ and $\Delta^*_1(u)=\Delta^*_2(u)=0$, one can truncate the flow equations by considering only $\Delta_{2,k}(u)=\lambda_{2,k}\cos(2u)$ since all harmonics of $\Delta_{1,k}(u)=\sum_{n\geq 1}\delta_{1,k}\cos(2nu)$ are irrelevant: 
\beq 
\begin{split}
	\dt \lambda_{2,k} &= (-3+\sig + 4 K_k l_1) \lambda_{2,k} + 4 (\bar L_1 - 2 \bar L_x) \lambda_{2,k}^2 , \\ 
	\dt K_k &= - 2 (\bar L_x + \bar L_\tau) \lambda_{2,k} K_k, \\ 
	\dt v_k &= 2 (\bar L_x - \bar L_\tau) \lambda_{2,k} v_k .
\end{split}
\label{floweq3} 
\eeq 
The relevant variables for the study of the nearly critical trajectories are $x=K_k-3/2+\sig/2$ and $y=\sqrt{\lambda_{2,k}}$. To obtain the flow equations to second order in these variables, it is sufficient to evaluate the threshold functions $l_1$, $\bar L_x$ and $\bar L_\tau$ for $\theta_{k}=0$ and $z_{k}=1$, which gives the universal value $l_1=1/2$ (independent of the choice of the function $r$ in~(\ref{Rdef})). Note that the term of order $\lambda_{2,k}^2$ in the first of Eqs.~(\ref{floweq3}) does not contribute to leading order. This leads to the standard equations of the BKT transition~\cite{Kosterlitz74sm,Chaikin_booksm}, 
\beq 
\begin{split}
	\dt x &= - (3-\sig) (\bar L_x + \bar L_\tau) y^2 , \\ 
	\dt y &= xy ,
\end{split}
\eeq 
while the velocity satisfies the (independent) equation
\beq 
\dt v_k = 2 (\bar L_x - \bar L_\tau) y^2 v_k . 
\eeq 
The critical trajectory in the plane $(x,y)$ corresponds to 
\beq 
y=\pm x [(3-\sig)|\bar L_x + \bar L_\tau|]^{-1/2} , 
\eeq 
in very good agreement with the numerical solution of the flow equations (see Fig.~\ref{fig_flowBKT} in the main text).

\section{II. Perturbative functional renormalization group} 

It has been pointed out that the perturbative (field-theoretical) FRG approach (PFRG) and the nonperturbative FRG (NPFRG) differ in one important respect when it comes to describing the Bose-glass phase induced by a short-range correlated disorder~\cite{Dupuis20sm}. While the NPFRG predicts the vanishing of the renormalized Luttinger parameter $K_k$ and the cusp in the renormalized functional disorder correlator $\Delta_k(u)$ to appear at infinite length scale ---at finite length scale the cusp in $\Delta_k(u)$ is rounded into a boundary layer whose size is controlled by $K_k$---, the PFRG predicts the vanishing of $K_k$ and the cusp in $\Delta_k(u)$ to occur at a finite scale. Yet the fixed point describing the Bose-glass phase is similar in both approaches, being characterized by a vanishing Luttinger parameter and a cusp in the functional disorder correlator. In this section, we obtain similar results in the case of a disorder with long-range correlations. While the vanishing of $K_k$ and the cusp in $\Delta_{1,k}(u)$ occur at finite length scale ($\Delta_{2,k}(u)$ remains analytic), the PFRG predicts two nontrivial fixed points, one of which corresponds to a Mott-glass phase. 

In order to apply the PFRG we generalize the Euclidean action, defined by  Eq.~(\ref{action}) in the main text, to $d$ space dimensions,
\begin{eqnarray}
	S[\{\phi_a\}] &=& \frac12 \sum\limits_{a=1}^{n} \int d^dx \int\limits_0^{\infty} d\tau \left[ Z_x (\dx\phi_a)^2 + Z_\tau (\dtau\phi_a)^2 \right] 
	-\frac12 \sum\limits_{a,b=1}^{n} \int d^dx  \int\limits_0^{\infty} d\tau
	\int\limits_0^{\infty} d\tau' \,
	V_{1}(\phi_a(x,\tau)-\phi_b(x,\tau')) \nonumber \\
	&& -  \frac12 \sum\limits_{a,b=1}^{n} \int d^dx \int d^dx' \int\limits_0^{\infty} d\tau
	\int\limits_0^{\infty} d\tau' \, 
	f(x-x') V_{2}(\phi_a(x,\tau)-\phi_b(x',\tau')),
	\label{eq:action-perturbative} 
\end{eqnarray} 
where 
\begin{eqnarray}
	Z_x = \frac{v} {\pi K}, \qquad  Z_\tau =  \frac1{\pi K v}.
\end{eqnarray} 
Using a double expansion in $\varepsilon=4-d$ and $\delta=3-\sigma$, we derive the perturbative FRG flow equations to lowest nontrivial order,
\begin{eqnarray}
	\partial_{\ell} \Delta_{1,\ell}(u) &=&(\varepsilon-2\eta_{x,\ell}) \Delta_{1,\ell}(u) +\tilde{K}_\ell  \Delta_{1,\ell}''(u) - \frac12 \frac{d^2}{d u^2}[\Delta_{1,\ell}(u)+\Delta_{2,\ell}(u)]^2
	+ [\Delta_{1,\ell}(0) +\Delta_{2,\ell}(0)]\Delta_{1,\ell}''(u), \label{eq:pertur-FRG-flow-1} \\
	\partial_{\ell} \Delta_{2,\ell}(u) &=& (\delta-2\eta_{x,\ell})\Delta_{2,\ell}(u) +
	\tilde{K}_\ell \Delta_{2,\ell}''(u)
	+ [\Delta_{1,\ell}(0) +\Delta_{2,\ell}(0)] \Delta_{2,\ell}''(u), \label{eq:pertur-FRG-flow-2} \\
	\partial_\ell \ln  \tilde{K}_\ell  &=&   1-d -\frac12 (\eta_{x,\ell}+\eta_{\tau,\ell}), \label{eq:flow-K} \\
	\partial_\ell \ln Z_{x,\ell} &=& \eta_{x,\ell}  = -\frac14 (\delta-\varepsilon) \Delta_{2,\ell}''(0) ,\\
	\partial_\ell \ln Z_{\tau,\ell} &=&  \eta_{\tau,\ell} =  -\Delta_{1,\ell}''(0) - \Delta_{2,\ell}''(0),
\end{eqnarray}
where we have introduced the dimensionless parameters
\beq 
\begin{split}
	\Delta_{1,\ell}(u) &= - \frac{\mathcal{A}_d V''_{1,\ell}(u)}{Z_{x,\ell}^2 \Lambda_\ell^{\varepsilon}},  \;\;\;
	\Delta_{2,\ell}(u) = - \frac{\mathcal{A}_d V''_{2,\ell}(u)}{Z_{x,\ell}^2\Lambda_\ell^{\delta}},  \;\;\;
	\tilde{K}_\ell =  \frac{\mathcal{A}_d \Lambda_\ell^{d-1}}{2\sqrt{Z_{x,\ell} Z_{\tau,\ell}}}.
\end{split}
\label{eq:pertur-FRG_flow-6} 
\eeq 
Here $\Lambda_\ell = \Lambda e^{-\ell}$ (with $\ell\geq 0$), $\mathcal{A}_d=S_d/(2\pi)^d$ and 
$S_d=2\pi^{d/2}/\Gamma(d/2)$ is the surface area of a  unit sphere in $d$-dimensional space. Note that for $d=1$ $\tilde{K}_{\ell=0}=K/2$.

Equations~(\ref{eq:pertur-FRG-flow-1}-\ref{eq:pertur-FRG_flow-6}) are similar to those obtained in the NPFRG approach if we set $d=1$, the only difference being that the threshold functions do not appear explicitly, and
$\eta_{x,\ell}$ and $\eta_{\tau,\ell}$ are obtained to lowest order in $\Delta_{1,\ell}$ and $\Delta_{2,\ell}$. This is the reason why, contrary to the NPFRG, the renormalized Luttinger parameter $\tilde{K}_\ell$ vanishes at a finite length scale for a small enough initial value [Eq.~(\ref{eq:flow-K})]. 
Indeed a finite $\tilde{K}_\ell>0$ rounds off the cusp appearing in $\Delta_{1,\ell}(u)$,
so that $\Delta_{1,\ell}(u)$ significantly  deviates form the fixed-point solution in the boundary layer $|u| \lesssim \tilde{K}_\ell$ and obeys the  following scaling form,
\begin{equation} \label{eq:layer}
	\Delta_{1,\ell}(u)=\Delta_{1,\ell}(0)+\tilde{K}_\ell \left[1-\sqrt{1+(u\chi/\tilde{K}_\ell)^2} \right],
\end{equation}
where $\chi=|\Delta_1^{*\prime}(0+)|$. Substituting the boundary-layer scaling (\ref{eq:layer})
in the flow equation~(\ref{eq:flow-K}), we obtain
\begin{equation} \label{eq:K-l}
	\partial_{\ell} \tilde{K}_\ell = -\theta_K\tilde{K}_\ell  - \frac{ \chi^2}{2},
\end{equation}
where $\theta_K=\frac14 (2+\delta-\varepsilon)  |\Delta_2^{*\prime\prime}(0)|$.
As follows from Eq.~(\ref{eq:K-l}), $\tilde{K}_\ell$ vanishes at a
finite length scale $\Lambda^{-1}e^{\ell_{\mathrm{loc}}}$, 
\begin{equation} \label{eq:l-loc}
	\ell_{\mathrm{loc}} =  \frac1{\theta_K}\ln\left(1 + \frac{2\theta_K \tilde{K}_0}{\chi^2} \right),
\end{equation}
at which the localization effects are set in.
While $\Delta_{2,\ell}(u)$ remains analytic along the FRG, as in the nonperturbative approach, $\Delta_{1,\ell}''(0)$ and $Z_{\tau,\ell}$ diverge at the finite length scale $\ell_{\mathrm{loc}}$. Beyond this length scale $\Delta_{1,\ell}(u)$ develops a cusp at the origin $u=0$ (and more generally at $u=p\pi$, $p\in\mathbb{Z}$) so that $\Delta_{1,\ell}''(0+)$ becomes positive and the flow approaches one of the two nontrivial ``zero-$K$'' fixed points: the ``short-range'' fixed point 
\begin{eqnarray}
	\textrm{(SR):} \ \ \  \tilde{K}^* &=&0, \ \ \ \Delta_1^* (u) = \frac{ \varepsilon}{6} \left[ \left(u-\frac{\pi}{2}\right)^2 - \frac{\pi^2}{12} \right], \ \ \ \Delta_2^*(u)=0, 
\end{eqnarray}
and the ``long-range'' fixed point 
\begin{eqnarray}
	\textrm{(LR):} \ \ \ \tilde{K}^* &=0&, \ \ \ \Delta_1^*(u) \neq 0   , \ \ \ \ \Delta_2^*(u)= \Delta_2^*(0) \cos(2u), \ \ \ \Delta_1^*(0)+\Delta_2^*(0) =\frac{\delta}{4}. 
\end{eqnarray}
These two fixed points are similar to those obtained in the NPFRG approach, and the condition of stability of the long-range fixed point, 
\begin{equation}
	\frac{\delta}{\varepsilon}>\frac{\pi^2}{9}\approx 1.097 ,
\end{equation}
coincides with the condition derived from Eq.~(\ref{eq:stability}) in the main text if we set $d=1$. Moreover we find that the superfluid stiffness $\rho_{s,\ell}=1/\pi^2 Z_{\tau,\ell}\sim \Lambda_\ell^{\eta_\tau^*}$ vanishes at both fixed points (since $\eta_\tau^*>0$), while the compressibility $\kappa_{\ell}=1/\pi^2 Z_{x,\ell}\sim \Lambda_\ell^{\eta^*_x}$ remains finite at the short-range fixed point ($\eta_x^*=0$) but vanishes at the long-range fixed point ($\eta_x^*>0$).

\section{III. The limit $\sig=-1$ and $K\to 0^+$}

In this section we consider the case $\sig=-1$ where the random potential $\xi(x)=\xi$ is perfectly correlated in both imaginary-time and space dimensions. For a given realization of the disorder, i.e. for a given value of $\xi=|\xi|e^{i\alpha}$, the action reads 
\begin{align}
	S &= \int_{x,\tau} \llbrace \frac{v}{2\pi K} \left[ (\dx\varphi)^2 + \frac{(\dtau\varphi)^2}{v^2} \right] + (\xi^* e^{2i\varphi} + \cc ) \rrbrace  \nonumber \\  
	&= \int_{x,\tau} \llbrace \frac{v}{2\pi K} \left[ (\dx\varphi)^2 + \frac{(\dtau\varphi)^2}{v^2} \right] - 2|\xi| \cos(2\varphi) \rrbrace ,
\end{align}
where we have shifted the field $\varphi(x,\tau)\to \varphi(x,\tau)+(\alpha+\pi)/2$. 
This action corresponds to a sine-Gordon model with a random coupling constant $|\xi|$. The BKT point for the model is located at $K_c=2$, which agrees with the result $K_c(\sig)=3/2-\sig/2$ obtained for an arbitrary $\sigma<0$. In the semiclassical limit $K\to 0^+$, one can expand about one of the minima of the potential, e.g. $\varphi=0$, which gives 
\begin{align}
	S &= \int_{x,\tau} \llbrace \frac{v}{2\pi K} \left[ (\dx\varphi)^2 + \frac{(\dtau\varphi)^2}{v^2} \right] + 4 |\xi|\varphi^2 \rrbrace \nonumber\\ 
	&= \int_{x,\tau} \frac{1}{2\pi Kv} \bigl[ v^2 (\dx\varphi)^2 + (\dtau\varphi)^2 + m^2 \varphi^2 \bigr] ,
\end{align}  
where $m=(8\pi vK|\xi|)^{1/2}$ is a random mass. The corresponding propagator reads 
\beq 
G(q,i\wn) = \mean{\varphi(q,i\wn) \varphi(-q,-i\wn)} = \frac{\pi Kv}{\wn^2+v^2 q^2 +m^2} ,
\label{Gdef} 
\eeq 
with $\wn=2n\pi T$ ($n\in\mathbb{Z}$) a Matsubara frequency.

\subsection{A. Compressibility and conductivity} 

The compressibility $\kappa_\xi=\partial\rho_0/\partial\mu$ can be obtained from  
\beq 
\kappa_\xi = \lim_{q\to 0} \chi_{\rho\rho}(q,0) = \lim_{q\to 0} \frac{q^2}{\pi^2} G(q,0) ,
\eeq 
where $\chi_{\rho\rho}(q,i\wn)=(q/\pi)^2 G(q,i\wn)$ is the long-wavelength part of the density-density correlation function. From~(\ref{Gdef}), we deduce 
\beq
\kappa_\xi = \llbrace 
\begin{array}{lll} 0 & \mbox{if} & \xi\neq 0 , \\ 
	\frac{K}{\pi v} & \mbox{if} & \xi=0 . 
\end{array}
\right. 
\eeq 
We conclude that the disorder-averaged compressibility $\kappa=\overline{\kappa_\xi}$ vanishes: The system is incompressible. 

On the other hand, the conductivity can be obtained from the density-density correlation function using gauge invariance~\cite{NDbook1sm}, 
\begin{align}
	\sig_\xi(\w) &= \lim_{q\to 0} \frac{\w+i0^+}{iq^2} \chi_{\rho\rho}(q,\w+i0^+) \nonumber\\ 
	&= i \frac{vK}{\pi} \frac{\w+i0^+}{(\w+i0^+)^2-m^2} .
\end{align}
Its real part is given by 
\beq 
\Re[\sig_\xi(\w)] = \frac{vK}{2} [ \delta(\w-m) + \delta(\w+m)] . 
\eeq 
The averaging over disorder must be done with the probability distribution function 
\beq 
P(\xi^*,\xi) = \frac{e^{-|\xi|^2/D}}{D} ,
\eeq 
which gives 
\beq 
\Re[\sig(\w)] = \int \frac{d\xi^* d\xi}{2i\pi} P(\xi^*,\xi)  \Re[\sig_\xi(\w)] 
= \frac{|\w|^3}{32\pi^2 KvD} \exp\left( - \frac{\w^4}{(8\pi Kv)^2 D} \right) 
\eeq 
using $(1/2i\pi) \int d\xi^*d\xi= (1/\pi) \int_0^{2\pi}d\alpha\int_0^\infty d|\xi|\, |\xi|$.  One easily verifies that the sum rule 
\beq 
\intinf d\w \, \Re[\sig(\w)] = vK 
\eeq 
is satisfied. The disorder-averaged conductivity of the incompressible system being gapless, the ground state is a Mott glass.

\subsection{B. Cumulants of the random functional $W[J;\xi]$} 

The random functional $W[J;\xi]=\ln Z[J;\xi]$ can be computed exactly since the action is Gaussian, 
\begin{align} 
	W[J;\xi] &= \ln \int \calD[\varphi] \, e^{-S[\varphi]+\int_{x,\tau}J\varphi} 
	\nonumber\\ 
	&= W[0;\xi] + \half \sum_{q,\wn} J(-q,-i\wn) G(q,i\wn) J(q,i\wn) , 
\end{align}
where $J(x,\tau)$ is an external source that couples linearly to the field $\varphi$, and 
\beq 
W[0;\xi] = - \half \sum_{q,\wn} \ln\left( \frac{\wn^2 + \w_q^2}{\pi K v} \right) 
\eeq 
with $\w_q=\sqrt{v^2q^2+m^2}$. The Matsubara sum is formally divergent. Since what matters is the dependence on the random mass $m$, we can subtract an infinite constant and consider 
\begin{align} 
	W[0;\xi] &= - \half \sum_{q,\wn} [ \ln(\wn^2 + \w_q^2) - \ln(\wn^2+v^2q^2) ] \nonumber\\ 
	&= - \frac{\beta}{2} \sum_q (\w_q - v|q|) ,
\end{align}
where the last result holds in the zero-temperature limit $\beta\to\infty$. The term $v|q|$ being independent of the random mass, we drop it in the following. We thus obtain 
\beq 
W[0;\xi] = - \beta L \left[ \frac{v\Lambda^2}{4\pi} + \frac{m^2}{4\pi v} \ln\left(\frac{2v\Lambda}{m}\right) \right] 
\eeq 
in the limit $m\ll v\Lambda$, where $\Lambda$ is the UV momentum cutoff of theory. For a uniform and time-independent source $J(x,\tau)=J$, i.e. $J(q,i\wn)=\sqrt{\beta L} \delta_{q,0}\delta_{\wn,0}J$, we finally find 
\beq 
W[J;\xi] = \beta L \left[ \frac{\pi Kv}{2m^2} J^2 - \frac{m^2}{4\pi v} \ln\left(   \frac{2v\Lambda}{m}\right) \right] 
\eeq 
dropping a constant term.

We are now in a position to calculate the cumulants of the random functional $W[J;\xi]$. The first cumulant is easily obtained, 
\beq 
W_1[J_a] = \overline{W[J_a;\xi]} = \beta L \left[ \frac{1}{16} \sqrt{\frac{\pi}{D}} J_a^2 - \frac{K}{2} \sqrt{\pi D} \ln \left( \frac{e^{-1+C/2}v\Lambda^2}{\pi K\sqrt{D}} \right) \right] , 
\eeq 
where $C\simeq 0.5772$ is Euler's constant. The second cumulant is given by 
\beq 
W_2[J_a,J_b] = \overline{W[J_a;\xi]W[J_b;\xi]} - W_1[J_a] W_1[J_b] .
\eeq 
The coefficient of the term $J_a^2J_b^2$ involves the average of $1/m^4$, 
\beq 
\overline{\left(\frac{1}{m^4}\right)} = \frac{2}{(8\pi Kv)^2D} \int_0^\infty d|\xi| \frac{e^{-|\xi|^2/D}}{|\xi|} , 
\eeq 
and thus diverges. Since $\Gamma_2[\phi_a,\phi_b]$ can be identified with $W_2[J_a,J_b]$, with an appropriate choice of the sources $J_a$ and $J_b$~\cite{Dupuis20sm}, we conclude that neither $W_2[J_a,J_b]$ nor $\Gamma_2[\phi_a,\phi_b]$ exist: The cumulant expansion breaks down when $\sigma=-1$.

\section{IV. Pinning of vortices by correlated disorder in type II superconductors} 

Here  we  want to elucidate an interesting application of our results to the physics of type-II superconductors. In these materials pinning of vortices by disorder plays a significant role in their dynamics and controls the critical current $J_c$. For this reason one can introduce disorder artificially, e.g.  irradiation by electrons or light ions can  produce numerous point-like defects. A more effective pinning can be realized by extended defects. For instance, irradiation of high-energy heavy ions can be used to create columnar defects and thus significantly enhance the pinning strength and  increase the critical current $J_c$.  

When the columnar defects are uncorrelated in the transverse direction, flux lines map onto bosons with a short-range correlated disorder~\cite{NeslonVinokur93sm}. This leads to a ``superfluid'' flux liquid at high temperatures  and a low-temperature Bose-glass  phase, in which the flux lines are localized on the columnar defects.  When the columnar disorder is algebraically correlated in the transverse direction, flux lines in a $(d+1)$-dimensional superconductor map onto $d$-dimensional bosons with long-range correlated disorder as described by Eq.~(\ref{eq:action-perturbative}). As we have shown the algebraically correlated columnar disorder may lead to a new phase, the Mott glass, which is incompressible contrary to the Bose glass.

In the flux line picture, $Z_x$ and $Z_\tau$ in the action~(\ref{eq:action-perturbative}) stand  for the compressional and tilt moduli, $c_{11}$ and $c_{44}$, respectively, while $K$ plays the role of temperature. The divergence of the tilt modulus $c_{44}$ leads to the transverse Meissner effect in both the Bose glass and  the  Mott glass: the vortex lines remain pinned to the disorder columns even though the magnetic field is tilted away from the column direction.  Applying  a magnetic field $H_{\bot}$ transverse to the direction of flux lines generates a new term, $| \partial_\tau \phi |$, in the action~\cite{Balents96sm}, similarly to the emergence of the threshold force term in the FRG flow for a driven elastic system at the depinning transition~\cite{Fedorenko06sm}. This also implies the appearance of a threshold $H_{\bot}^c$ in the response $B_{\bot}$ of the system to the tilt of flux lines by the transverse  magnetic field $H_{\bot}$ such that the response has a depinning form~\cite{Fedorenko:2008sm}  
\begin{equation}
	B_{\bot}\sim(H_{\bot}-H_{\bot}^c)^{\phi},
\end{equation}
where $B_{\bot}$ is the transverse magnetic induction due to the
tilted flux lines. The critical exponent 
\begin{equation}
	\phi=(1+\Delta_1^{*\prime\prime}(0+)+\Delta_2^{*\prime\prime}(0+))^{-1}
\end{equation}
takes different universal  values in the Bose-glass and  Mott-glass phases.
Since the compressional modulus $c_{11}$ diverges  at the long-range fixed point and remains finite at the short-range fixed point, contrary to the Bose glass the Mott glass is incompressible --  the vortex density and thus the magnetic induction  $B_\parallel$ remain constant over a finite range of external magnetic field $H_\parallel$. Thus the Mott glass behaves in many respects like a Meissner phase but at a finite magnetic induction $B_\parallel$ instead of $B_\parallel=0$.  A similar behavior was predicted  for the Mott-insulator phase~\cite{NeslonVinokur93sm} which was realised in a superconductor with periodic holes on the surface at matching magnetic fields~ \cite{Goldberg:2009sm}. The difference between the Mott-insulator and Mott-glass phases lies in the spectrum of the vibrational modes of the vortex lattice. In the Bose/Mott glass the spectrum is gapless, whereas it is gapped in the Mott insulator.

\end{document}